\newcommand{\BR}{{\cal B}}
\newcommand{\y}{Y(4260)}

\newcommand{\xx}{X(3872)}
\newcommand{\pp}{\pi^+\pi^-}
\newcommand{\ks}{K_S^0}
\newcommand{\pip}{\pi^+}
\newcommand{\pim}{\pi^-}

\newcommand{\LL}{\ell^+\ell^-}
\newcommand{\EE}{e^+e^-}
\newcommand{\MM}{\mu^+\mu^-}

\newcommand{\hc}{h_c}
\newcommand{\pphc}{\pi^+\pi^- h_c}

\newcommand{\psip}{\psi(2S)}
\newcommand{\pspp}{\psi(3770)}
\newcommand{\jpsi}{J/\psi}
\newcommand{\piz}{\pi^0}

\newcommand{\zc}{Z_c(3900)}
\newcommand{\zcp}{Z_c(4020)}

\newcommand{\ppjpsi}{\pi^+\pi^-J/\psi}
\def\Journal#1#2#3#4{{#1} {\bf #2}, #3 (#4)}

\def\PLB{Phys. Lett. B}
\def\PRL{Phys. Rev. Lett.}
\def\PRD{Phys. Rev. D}

\def\EPJC{Eur. Phys. J. C}

\documentclass{ws-ijmpa}
\usepackage{cite}
\usepackage{graphicx}
\begin{document}
\markboth{Authors' Names} {Instructions for Typing Manuscripts
(Paper's Title)}

\catchline{}{}{}{}{}
%
%%%%%%%%%%%%%%%%%%%%%%%%%%%%%%%%%%%%%%%%%%%%%%%%%%%%%%%%%%%%%%%%%%%%

\title{Recent progress on the study of the charmoniumlike
states\footnote{Proceedings of the invited talk at the XXVI
International Symposium on Lepton Photon Interactions at High
Energies, June 2013, San Francisco, USA. The results are not
limited to those before June 2013.}}

\author{Chang-Zheng Yuan}

\address{Institute of High Energy Physics, Chinese Academy of
Sciences, \\ 19B Yuquan Road, Beijing 100049, China\\
yuancz@ihep.ac.cn}

\maketitle

\begin{history}
\received{Day Month Year}
\revised{Day Month Year}
\end{history}

\begin{abstract}

In this article, we review the recent experimental studies on the
charmoniumlike states, mainly from the $\EE$ annihilation
experiments BESIII, Belle, BaBar, and CLEO-c, and the hadron
collider experiment LHCb. We discuss the results on the $X(3872)$,
the vector $Y$ states [$Y(4008)$, $Y(4660)$, and those in $\EE\to
\pphc$], and the charged charmoniumlike $Z_c^-$ states.

\end{abstract}

\ccode{PACS numbers: 14.40.Rt, 14.40.Pq, 13.66.Bc}

%\tableofcontents

\section{Introduction}

In the conventional quark model, mesons are composed of one quark
and one anti-quark, while baryons are composed of three
quarks~\cite{GellMann}. However, exotic hadronic states with other
configurations have been searched for and many candidates were
proposed~\cite{klempt,epjc-review}. These states include glueballs
(with no quark), hybrids (with quarks and excited gluon),
multi-quark states (with more than three quarks), and hadron
molecules (bound state of two or more hadrons). Since a proton and
a neutron can be bounded to form a deuteron, it is also believed
that other states beyond the quark model must exist.

It is a long history of searching for all these kinds of states,
however, no solid conclusion was reached until recently on the
existence of any one of them, except
deuteron~\cite{klempt,epjc-review}. Two statements may well
describe the situation as of 2005 when the existence of the
$\Theta(1540)$, a candidate of pentaquark state was
discussed~\cite{theta1540_review}: ``The absence of exotics is one
of the most obvious features of QCD''~\cite{jaffe_2005}, and ``The
story of pentaquark shows how poorly we understand
QCD''~\cite{wilczek_2005}.

Dramatic progress was made in the study of the exotic states after
the running of the two $B$-factories, i.e., Belle~\cite{belle} at
KEK and BaBar~\cite{babar} at SLAC. Although the two experiments
were designed for the study of the $B$ mesons, copious events with
charm and anti-charm quark pair were produced due to the
unprecedented high luminosity reached at these two experiments.
This made the study of the charmonium spectroscopy very fruitful,
and since 2003, lots of new states (called charmoniumlike states
or XYZ particles) have been observed in the final states with a
charmonium and some light
hadrons~\cite{epjc-review,QWG_YR,swanson,godfrey}.

All these states populate in the charmonium mass region. They
could be candidates for charmonium states, however, there are also
strange properties shown from these states, these make them more
like exotic states rather than conventional
mesons~\cite{epjc-review,swanson,godfrey}.

The BESIII~\cite{bes3} experiment at the BEPCII collider started
data taking in 2009, and lots of data were accumulated at the peak
of the narrow vector charmonium resonances like $\jpsi$, $\psip$,
and $\pspp$, as well as at above 4~GeV, these high energy data
make the study of the XYZ states possible~\cite{zc4020}.

The LHCb~\cite{lhcb} experiment at the LHC accumulated 3~fb$^{-1}$
$pp$ collision data at $\sqrt{s} = 7$ and 8~TeV, corresponding to
more than an order of magnitude more $B$ decays compared with the
BaBar and Belle experiments. This data sample made the improved
study of the XYZ states observed in B decays
possible~\cite{LHCbx,LHCbgammapsp,LHCbzc4430}.

In this review, we present the most recent results on the study of
the $X(3872)$ from the $B$ decays and the $\y$ radiative
transition, the $Y$ states from initial state radiation (ISR)
processes and from $\EE$ annihilation, and the charged $Z_c^-$
states. The results are from the BESIII~\cite{bes3},
Belle~\cite{belle}, BaBar~\cite{babar}, and CLEO-c~\cite{cleoc}
experiments, as well as the LHCb~\cite{lhcb}. The data samples
used for the study of the XYZ states collected at these
experiments are summarized in Table~\ref{datasample}.

\begin{table}[htbp]
\tbl{Data samples for the study of the charmoniumlike states.}
{\begin{tabular}{@{}crrc@{}} \toprule
  Experiment & $\sqrt{s}$~(GeV) & ${\cal L}$ (fb$^{-1}$) & comments  \\
  \hline
  BaBar    & 10.023  &   14~~~    &  $\Upsilon(2S)$ \\
           & 10.355  &   30~~~    &  $\Upsilon(3S)$ \\
           & 10.580  &  433~~~    &  $\Upsilon(4S)$ \\
           & 10.520  &   54~~~    &  off resonance      \\ \hline
  Belle    & 9.460   &    6~~~    &  $\Upsilon(1S)$ \\
           & 10.023  &   25~~~    &  $\Upsilon(2S)$ \\
           & 10.355  &    3~~~    &  $\Upsilon(3S)$ \\
           & 10.580  &  702~~~    &  $\Upsilon(4S)$ \\
           & 10.867  &  121~~~    &  $\Upsilon(5S)$ \\
           & 10.520  &   89~~~    &  off resonance \\
      & 10.75-11.02  &   28~~~    &  $\Upsilon(5S)$ scan \\ \hline
  BESIII   &  4.009  & 0.5~~~     &  $\psi(4040)$\\
           &  4.230  & 1.1~~~     &  $Y(4260)$  \\
           &  4.260  & 0.8~~~     &  $Y(4260)$   \\
           &  4.360  & 0.5~~~     &  $Y(4360)$   \\
           &  4.420  & 1.0~~~     &  $\psi(4415)$ \\
           &  4.600  & 0.6~~~     &  $Y(4660)$  \\
         & 3.81-4.59 & 1.6~~~     &  R scan and off resonance \\\hline
  CLEO-c   &  4.170  & 0.6~~~     &  $\psi(4160)$  \\\hline
    LHCb   & $7\times 10^3$  & 1~~~ & $pp$ collision \\
           & $8\times 10^3$  & 2~~~ & $pp$ collision \\ \botrule
\end{tabular}\label{datasample}}
\end{table}

\section{\boldmath New information on the $\xx$}

The $\xx$ was observed at the Belle experiment in $B^\pm\to K^\pm
\ppjpsi$ decays more than ten years ago~\cite{bellex}. It was
confirmed subsequently by several other
experiments~\cite{CDFx,D0x,babarx}. Since its discovery, the $\xx$
state has stimulated special interest for its nature. Both BaBar
and Belle observed $\xx\to \gamma\jpsi$ decay, which supports
$\xx$ being a C-even state~\cite{babar-jpc,belle-jpc}. The CDF and
LHCb experiments determined the spin-parity of the $\xx$ being
$J^{P}=1^{+}$~\cite{CDF-jpc,LHCbx}, and CDF experiment also found
that the $\pp$ system was dominated by a $\rho^0(770)$ resonance
in $\xx\to \ppjpsi$~\cite{CDF-pp}.

The $\xx$ was only observed in $B$ meson decays and hadron
collisions before. Since the quantum number of $\xx$ is
$J^{PC}=1^{++}$, it can be produced through the radiative
transition of the excited vector charmonium or charmoniumlike
states such as the $\psi$s and the $Y$s~\cite{BES3x}. Being near
$D^0\bar{D}^{*0}$ mass threshold, the $\xx$ was interpreted as a
good candidate for a hadronic molecule or a tetraquark
state~\cite{epjc-review}, the observation of its large decay rates
to $\gamma\psip$ contradicts expectation of the molecule
model~\cite{babar_x_gpsp,LHCbgammapsp}. These will be discussed
below.

\subsection{Observation of $\y\to \gamma \xx$}

BESIII reported the observation of $\EE\to \gamma\xx\to \gamma
\ppjpsi$, with $\jpsi$ reconstructed through its decays into
lepton pairs ($\LL=\EE$ or $\MM$)~\cite{BES3x}. The analysis is
performed with the data samples collected with the BESIII detector
taken at $\EE$ center-of-mass (CM) energies from 4.009~GeV to
4.420~GeV~\cite{zc4020}.

\begin{figure}
\begin{center}
\includegraphics[height=7cm]{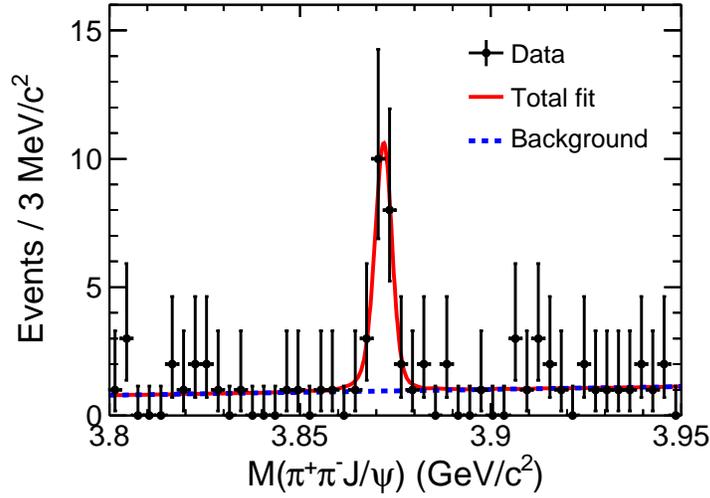}
\caption{Fit the $M(\ppjpsi)$ distribution observed at BESIII.
Dots with error bars are data, the curves are the best fit.}
\label{fit-mx}
\end{center}
\end{figure}

The $M(\ppjpsi)$ distribution (summed over all energy points), as
is shown in Fig.~\ref{fit-mx}, is fitted to extract the mass and
signal yield of the $\xx$. The ISR $\psip$ signal is used to
calibrate the absolute mass scale and to extract the resolution
difference between data and MC simulation. Figure~\ref{fit-mx}
shows the fit result, the measured mass of $\xx$ is $(3871.9\pm
0.7\pm 0.2)$~MeV/$c^2$. From a fit with a floating width one
obtains a width of $(0.0^{+1.7}_{-0.0})$~MeV, or less than 2.4~MeV
at the 90\% confidence level (C.L.). The statistical significance
of $\xx$ is $6.3\sigma$, estimated by comparing the difference of
log-likelihood value with and without $\xx$ signal in the fit, and
taking the change of number of degrees of freedom into
consideration.

The Born-order cross section is calculated using
 $
\sigma^{B}=\frac{N^{\rm obs}} {\mathcal{L}_{\rm int} (1+\delta)
\epsilon \mathcal{B}},
 $
where $N^{\rm obs}$ is the number of observed events obtained from
the fit to the $M(\ppjpsi)$ distribution, $\mathcal{L}_{\rm int}$
is integrated luminosity, $\epsilon$ is selection efficiency,
$\mathcal{B}$ is branching ratio of $\jpsi\to \LL$ and
($1+\delta$) is the radiative correction factor. The results are
listed in Table~\ref{sec}. For 4.009~GeV and 4.360~GeV data, since
the $\xx$ signal is not significant, upper limits on the
production rates are given at 90\% C.L. The measured cross
sections at around 4.260~GeV are an order of magnitude higher than
the NRQCD calculation of continuum production~\cite{ktchao}, this
may suggest the $\xx$ events come from a resonant decays.

\begin{table}[htbp]
\tbl{The product of the Born cross section $\sigma^{B}(\EE\to
\gamma \xx)$ and $\mathcal{B}(\xx\to \ppjpsi)$ at different energy
points. The upper limits are given at 90\% C.L.}
{\begin{tabular}{@{}cc@{}} \toprule
  $\sqrt{s}$~(GeV)  & $\sigma^B[\EE\to \gamma\xx]
           \cdot\mathcal{B}(\xx\to \ppjpsi)$~(pb) \\  \hline
  4.009 &  $0.00\pm 0.04\pm 0.01$ or $<0.11$  \\
  4.229 &  $0.27\pm 0.09\pm 0.02$  \\
  4.260 &  $0.33\pm 0.12\pm 0.02$  \\
  4.360 &  $0.11\pm 0.09\pm 0.01$ or $<0.36$ \\ \botrule
\end{tabular}\label{sec}}
\end{table}

\begin{figure}
\begin{center}
\includegraphics[height=6cm]{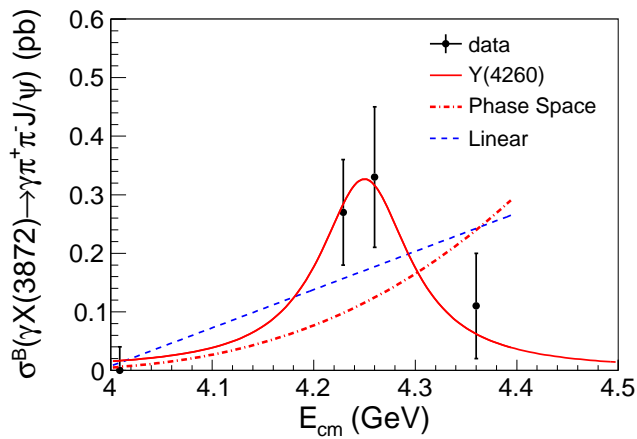}
\caption{The fit to $\sigma^B[\EE\to \gamma\xx]\times
\mathcal{B}[\xx\to \ppjpsi]$ measured by BESIII with a $\y$
resonance (red solid curve), a linear continuum (blue dashed
curve), or an $E1$-transition phase space term (red dotted-dashed
curve). Dots with error bars are data.} \label{xsec}
\end{center}
\end{figure}

The energy-dependent cross sections are fitted with a $\y$
resonance (parameters fixed to PDG~\cite{pdg} values), a linear
continuum, or an $E1$-transition phase space ($\propto
E^3_\gamma$) term. Figure~\ref{xsec} shows all the fit results,
which give $\chi^2/{\rm ndf}=0.49/3$ (C.L.=92\%), 5.5/2
(C.L.=6\%), and 8.7/3 (C.L.=3\%) for a $\y$ resonance, linear
continuum, and phase space distribution, respectively. The $\y$
resonance describes the data better than the other two options.

These observations strongly support the existence of the radiative
transition process $\y\to \gamma\xx$. The $\y\to \gamma\xx$ could
be another previously unseen decay mode of the $\y$ resonance.
This, together with the transitions to the charged charmoniumlike
state $\zc$~\cite{zc3900,belley_new,seth_zc}, suggest that there
might be some commonality in the nature of the $\xx$, $\y$, and
$\zc$, the model developed to interpret any one of them should
also consider the other two. As an example, the authors of
Ref.~\refcite{model} put all these states into a molecular picture
to calculate $\EE\to \gamma \xx$ cross sections.

Combining with the $\EE\to \ppjpsi$ cross section measurement at
$\sqrt{s}=4.260$~GeV from BESIII~\cite{zc3900}, one obtains
$\sigma^B[\EE\to \gamma\xx]\cdot \BR[\xx\to
\ppjpsi]/\sigma^B(\EE\to \ppjpsi) = (5.2\pm 1.9)\times 10^{-3}$,
under the assumption that $\xx$ and $\ppjpsi$ produced only from
$\y$ decays. If one takes $\BR[\xx\to \ppjpsi] =
5\%$~\cite{bnote}, then $\mathcal{R} = \frac{\BR[\y\to
\gamma\xx]}{\BR(\y\to \ppjpsi)}\sim 0.1$.

\subsection{Observation of $\xx\to \gamma \psip$}

Radiative decays of the $\xx$ provide crucial information to
understand its nature, especially to check if it is a conventional
charmonium or an exotic state.

The transition $\xx\to \gamma \jpsi$ was measured by
BaBar~\cite{babar-jpc} and Belle experiments~\cite{belle-jpc}.
Evidence for the $\xx\to \gamma \psip$ was reported by
BaBar~\cite{babar_x_gpsp} with a statistical significance of
$3.5\sigma$ and the ratio of the branching fractions was measured
to be
\[
R = \frac{\BR(\xx\to \gamma\psip)}{\BR(\xx\to \gamma\jpsi)} = 3.4
\pm 1.4.
\]
In contrast, no significant signal was observed at Belle and an
upper limit of $R < 2.1$ was reported at the 90\%
C.L.~\cite{belle-jpc} (using the information from
Ref.~\refcite{belle-jpc}, we can get \( R = \frac{\BR(\xx\to
\gamma\psip)}{\BR(\xx\to \gamma\jpsi)} = 0.6 \pm 1.4 \) as a good
estimation of the central value and uncertainty). Although not in
disagreement, BaBar and Belle results do show some tension on the
decay rate of $\xx\to \gamma\psip$.

In a recent study at the LHCb experiment, strong evidence for the
decay $\xx\to\gamma \psip$ was reported together with a
measurement of $R$~\cite{LHCbgammapsp}. The analysis is based on a
data sample of 1~fb$^{-1}$ at $\sqrt{s}=7$~TeV and 2~fb$^{-1}$ at
$\sqrt{s}=8$~TeV. In the full data sample, $591\pm 48$ $B^\pm \to
\xx K^\pm$, with $\xx\to \gamma \jpsi$, and $36.4\pm 9.0$ $B^\pm
\to \xx K^\pm$, with $\xx\to \gamma \psip$ were observed. The
significance of the $\xx\to \gamma \psip$ signal is determined by
simulating a large number of background-only toy MC experiments,
taking into account all the uncertainties in the shape of the
background distribution. The probability for the background to
fluctuate to at least the number of observed events is found to be
$1.2\times 10^{-5}$, corresponding to a significance of
$4.4\sigma$.

LHCb measures
\[
R = \frac{\BR(\xx\to \gamma\psip)}{\BR(\xx\to \gamma\jpsi)} =
2.46\pm 0.64\pm 0.29.
\]
This result is compatible with, but more precise than the BaBar
and Belle measurements~\cite{babar_x_gpsp,belle-jpc}.

As the measurements of all the above three experiments agree with
each other, we can make a weighted average to give the best
estimation of $R$. Neglecting the small correlated errors in the
measurements, we obtain
\[
\overline{R} = \frac{\BR(\xx\to \gamma\psip)}{\BR(\xx\to
\gamma\jpsi)} = 2.31\pm 0.57.
\]
This value does not support a pure $\bar{D^0}D^{*0}$ molecular
interpretation of the $\xx$ state, but agrees with expectations if
the $\xx$ is a pure charmonium or a mixture of a molecule and a
charmonium~\cite{epjc-review,LHCbgammapsp}. Of course many of the
calculations have model-dependent parameters, adjustment of the
parameters may still reproduce the experimental data.

\section{More information on the $Y$ states}

The study of charmonium states via ISR at the $B$-factories has
proven to be very fruitful. In the process $e^+e^- \to \gamma_{\rm
ISR} \pi^+\pi^-J/\psi$, the BaBar experiment observed the
$Y(4260)$~\cite{babary}. This structure was also observed by the
CLEO~\cite{cleoy} and Belle experiments~\cite{belley} with the
same technique; moreover, there is a broad structure near
4.008~GeV in the Belle data. In an analysis of the $e^+e^- \to
\gamma_{\rm ISR} \pi^+\pi^-\psi(2S)$ process, BaBar found a
structure at around 4.32~GeV~\cite{babar_pppsp}, while the Belle
observed two resonant structures at 4.36~GeV and
4.66~GeV~\cite{belle_pppsp}. Recently, BaBar updated $e^+e^- \to
\gamma_{\rm ISR} \pi^+\pi^-\psi(2S)$ analysis with the full data
sample, and confirmed the $Y(4360)$ and $Y(4660)$
states~\cite{babar_pppsp_new}; and the update with the full Belle
data samples further improve the measurements on the resonant
structures~\cite{belle_pppsp_new}. The update of the $e^+e^- \to
\gamma_{\rm ISR} \pi^+\pi^-J/\psi$ from both the BaBar and Belle
experiments still show differences at the $Y(4008)$ mass
region~\cite{babary_new,belley_new}.

BESIII experiment reported the cross section of $\EE\to \pp h_c$
final state with 13 energy points between 3.81 and
4.42~GeV~\cite{zc4020}, together with the CLEO-c measurement at
4.17~GeV~\cite{cleoc_pipihc}, the data indicate the existence of a
narrow structure at around 4.22~GeV and a wide structure at
4.29~GeV~\cite{y4220_ycz}.

\subsection{\boldmath Confirmation of the $Y(4660)$}

The BaBar experiment reported the update of the study of the
process $e^+e^-\to \pp\psi(2S)$ with ISR
events~\cite{babar_pppsp_new} with the full data sample. The data
were recorded with the BaBar detector at CM energies at and near
the $\Upsilon(nS)$ ($n$ = 2, 3, 4) resonances and correspond to an
integrated luminosity of 520~fb$^{-1}$. The $\psip$ is
reconstructed with its decays into either $\pp\jpsi$ or $\MM$.
They investigate the $\pp\psi(2S)$ mass distribution from 3.95 to
5.95~GeV, and measure the CM energy dependence of the associated
$e^+e^-\to \pp\psi(2S)$ cross section. The mass distribution
exhibits evidence for two resonant structures. A fit to the
$\pp\psi(2S)$ mass distribution corresponding to the decay mode
$\psi(2S)\to \pp J/\psi$ yields a mass value of $(4340\pm 16\pm
9)$~MeV/$c^2$ and a width of $(94\pm 32\pm 13)$~MeV for the
$Y(4360)$, and for the $Y(4660)$ a mass value of $(4669\pm 21\pm
3)$~MeV/$c^2$ and a width of $(104\pm 48\pm
10)$~MeV~\cite{babar_pppsp_new}. The results are in good agreement
with the Belle measurement~\cite{belle_pppsp} and confirm the
$Y(4660)$ observed by the Belle experiment.

Using the 980~fb$^{-1}$ full data sample taken with the Belle
detector, Belle also updated the analysis with two $\psip$ decay
modes~\cite{belle_pppsp_new}, namely, $\pp\jpsi$ and $\MM$.

\begin{figure}[htbp]
\centering
 \psfig{file=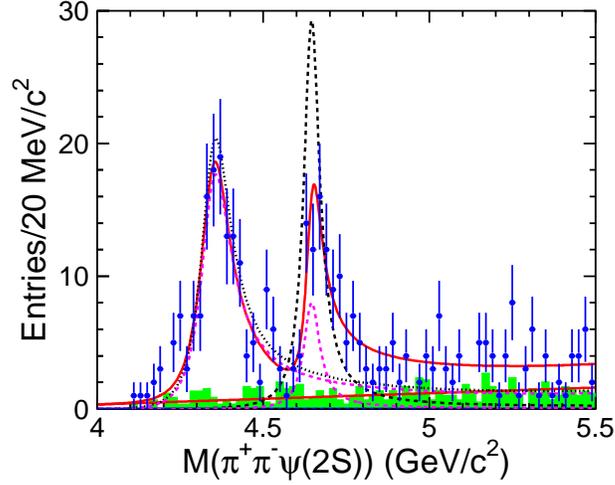, height=8.0cm, angle=-90}
\caption{The $\pp\psip$ invariant mass distribution from the Belle
experiment and the fit results with the coherent sum of two BW
functions. The sum of $\pp\jpsi$ and $\MM$ modes is shown. The
points with error bars are data while the shaded histograms the
normalized $\psip$ sideband backgrounds. The curves show the best
fit and the dashed curves, which are from the two solutions, show
the contributions from different BW components. The interference
between the two resonances is not shown.} \label{2bwfit}
\end{figure}

Fitting the mass spectrum of $\pp\psip$ with two coherent BW
functions (see Fig.~\ref{2bwfit}), Belle obtains $M[Y(4360)] =
(4346\pm 6\pm 2)$~MeV/$c^2$, $\Gamma[Y(4360)] = (111\pm 10\pm
7)$~MeV, $M[Y(4660)] = (4644\pm 12\pm 8)$~MeV/$c^2$ and
$\Gamma[Y(4660)] = (59\pm 12\pm 2)$~MeV. There are two solutions
from the fit. One solution has $\BR[Y(4360)\to \pp\psip]\cdot
\Gamma^{Y(4360)}_{\EE} = (10.6\pm 0.6\pm 0.7)$~eV and
$\BR[Y(4660)\to \pp\psip]\cdot \Gamma^{Y(4660)}_{\EE} = (6.8\pm
1.6\pm 0.7)$~eV, while the other $\BR[Y(4360)\to \pp\psip]\cdot
\Gamma^{Y(4360)}_{\EE} = (9.2\pm 0.8\pm 0.7)$~eV and
$\BR[Y(4660)\to \pp\psip]\cdot \Gamma^{Y(4660)}_{\EE} = (1.8\pm
0.3\pm 0.1)$~eV.

\begin{figure}[htbp]
\centering
 \psfig{file=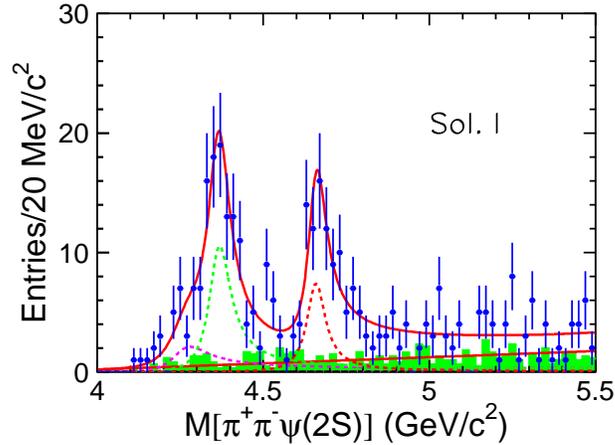, height=8.0cm, angle=-90}
\caption{Same as Fig.~\ref{2bwfit} but fit with the coherent sum
of three BW functions. } \label{fit-3r}
\end{figure}

Since there are some events accumulating at the mass region of
$Y(4260)$, the fit with the $Y(4260)$ included is also performed.
In the fit, the mass and width of the $Y(4260)$ are fixed to the
latest measured values at Belle~\cite{belley_new}. There are four
solutions with equally good fit quality. The signal significance
of the $Y(4260)$ is estimated to be $2.1\sigma$. The fit results
are shown in Fig.~\ref{fit-3r} for one of the solutions. In this
fit, one obtains $M[Y(4360)]=(4363\pm 8)$~MeV/$c^2$,
$\Gamma[Y(4360)]=(80\pm 16)$~MeV, $M[Y(4660)]=(4657\pm
9)$~MeV/$c^2$, and $\Gamma[Y(4660)]=(68\pm 11)$~MeV. Here the
errors are statistical only.

\begin{figure}[htbp]
\centering
 \psfig{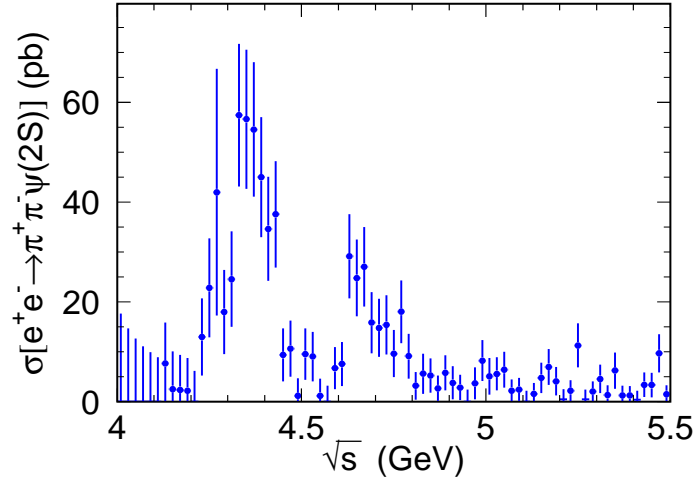}
\caption{The measured $\EE\to \pp\psip$ cross section for
$\sqrt{s}$=4.0 to 5.5~GeV from the Belle experiment. The errors
are the summed statistical errors of the numbers of signal and
background events. A common systematic error of 4.8\% for all the
data points is not shown.} \label{xs_full}
\end{figure}

The cross section for $\EE\to \pp\psip$ in each $\pp\psip$ mass
bin is calculated and the results in the full solid angle are
shown in Fig.~\ref{xs_full}, where the error bars include the
statistical uncertainties in the signal and the background
subtraction. The systematic error for the cross section
measurement is 4.8\% and common to all the data points.

\begin{figure}[htbp]
\centering
 \psfig{file=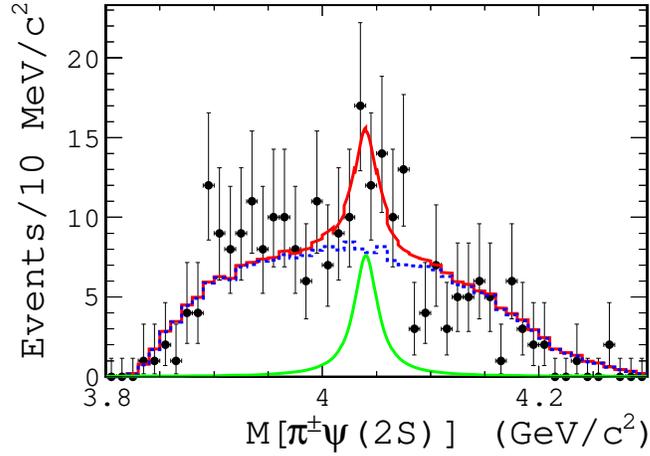,width=10.0cm}
\caption{The distribution of $M_{\pi^{\pm}\psip}$ from the sum of
the $\pp\jpsi$ and the $\MM$ modes for $Y(4360)$ events observed
in Belle data. The curve shows the best fit. This blank histogram
is the MC simulation of three-body phase space. }
\label{mppsp-fit}
\end{figure}

Possible charged charmoniumlike structures in $\pi^{\pm}\psip$
final states from the $Y(4360)$ or $Y(4660)$ decays are searched
for with the selected candidate events at Belle.
Figure~\ref{mppsp-fit} shows the sum of $M_{\pim\psip}$ and
$M_{\pip\psip}$ distributions in $Y(4360)$ decays from both the
$\ppjpsi$ and the $\MM$ modes. There is a bump at $4.05$~GeV/$c^2$
in the $\pi^\pm\psip$ invariant mass distribution, which could be
the $\zcp$~\cite{zc4020} or the $Z_c(4025)$~\cite{zc4025}. A
simple fit with a {\rm BW} function for the bump and a MC
simulated three-body phase space for the non-resonant background
yields a mass of $(4040\pm 9)$~MeV/$c^2$ and a width of $(26\pm
18)$~MeV. Here the errors are statistical only. The statistical
significance of the signal is $2.2\sigma$. Since each event is
counted twice, the significance is a bit overestimated. The same
distribution in $Y(4660)$ decays is also checked. The $Y(4660)$
sample is very limited in statistics, and there is no significant
structures in the $\pi^\pm\psip$ system.

\subsection{\boldmath Measurement of $\EE\to \pp h_c$}
\label{sec_y}

BESIII studied $\EE\to \pp h_c$ at 13 CM energies from 3.900 to
4.420~GeV~\cite{zc4020}. In the studies, the $h_c$ is
reconstructed via its electric-dipole (E1) transition $h_c\to
\gamma\eta_c$ with $\eta_c\to X_i$, where $X_i$ signifies 16
exclusive hadronic final states: $p \bar{p}$, $2(\pi^+ \pi^-)$,
$2(K^+ K^-)$, $K^+ K^- \pi^+ \pi^-$, $p \bar{p} \pi^+ \pi^-$,
$3(\pi^+ \pi^-)$, $K^+ K^- 2(\pi^+ \pi^-)$, $\ks K^\pm \pi^\mp$,
$\ks K^\pm \pi^\mp \pi^\pm \pi^\mp$, $K^+ K^- \pi^0$, $p
\bar{p}\pi^0$, $\pi^+ \pi^- \eta$, $K^+ K^- \eta$, $2(\pi^+ \pi^-)
\eta$, $\pi^+ \pi^- \pi^0 \pi^0$, and $2(\pi^+ \pi^-) \pi^0
\pi^0$.

\begin{figure}[htbp]
\centering
 \includegraphics[width=10cm]{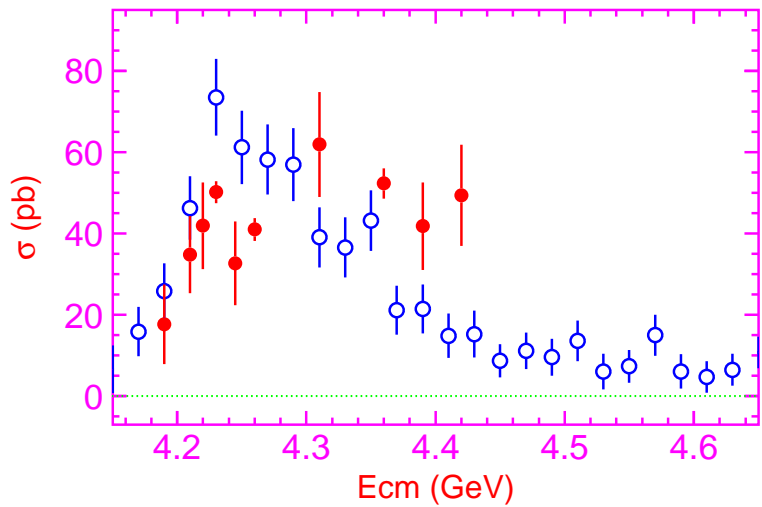}
\caption{The comparison between the cross sections of $\EE\to \pp
h_c$ from BESIII~(dots with error bars) and those of $\EE\to
\ppjpsi$ from Belle~(open circles with error bars). The errors are
statistical only.}
  \label{pphc_ppjpsi}
\end{figure}

The cross sections are listed in Table~\ref{scan-data} and shown
in Fig.~\ref{pphc_ppjpsi}. The CLEO-c experiment did a similar
analysis, but with significant signal only at CM energy
4.17~GeV~\cite{cleoc_pipihc}, the result is $\sigma=(15.6\pm
2.3\pm 1.9\pm 3.0)$~pb, where the third error is from the
uncertainty in $\BR[\psip\to \piz\hc]$.

\begin{table}[htbp]
\tbl{$\EE\to \pphc$ cross sections measured from the BESIII
experiment. For the first three energy points, besides the upper
limits, the central values and the statistical errors which will
be used in the fits below are also listed. The second errors are
systematic errors and the third ones are from the uncertainty in
$\BR(h_c\to \gamma\eta_c)$.}
 {\begin{tabular}{@{}cc@{}} \toprule
  $\sqrt{s}$~(GeV) &  $\sigma(\EE\to \pphc)$~(pb) \\
  \hline
  3.900    & $0.0\pm 6.0$ or $<8.3$ \\
  4.009    & $1.9\pm 1.9$ or $<5.0$ \\
  4.090    & $0.0\pm 7.4$ or $<13$ \\
  4.190    & $17.7\pm  9.8\pm  1.6\pm 2.8$ \\
  4.210    & $34.8\pm  9.5\pm  3.2\pm 5.5$ \\
  4.220    & $41.9\pm 10.7\pm  3.8\pm 6.6$ \\
  4.230    & $50.2\pm  2.7\pm  4.6\pm 7.9$ \\
  4.245    & $32.7\pm 10.3\pm  3.0\pm 5.1$ \\
  4.260    & $41.0\pm  2.8\pm  3.7\pm 6.4$ \\
  4.310    & $61.9\pm 12.9\pm  5.6\pm 9.7$ \\
  4.360    & $52.3\pm  3.7\pm  4.8\pm 8.2$ \\
  4.390    & $41.8\pm 10.8\pm  3.8\pm 6.6$ \\
  4.420    & $49.4\pm 12.4\pm  4.5\pm 7.6$ \\
  \botrule
\end{tabular}\label{scan-data}}
\end{table}

The cross sections are of the same order of magnitude as those of
the $\EE\to \ppjpsi$ measured by previous
experiments~\cite{zc3900,babary_new,belley_new}, but with a
different line shape (see Fig.~\ref{pphc_ppjpsi}). There is a
broad structure at high energy with a possible local maximum at
around 4.23~GeV. The BESIII and the CLEO-c measurements are used
to extract the resonant structures in $\EE\to \pp\hc$ in
Ref.~\refcite{y4220_ycz}.

As the systematic error ($\pm 18.1\%$) of the BESIII experiment is
common for all the data points, only the statistical errors are
used in the fits. The CLEO-c measurement is independent from the
BESIII experiment, and all the errors added in quadrature ($\pm
4.2$~pb) is taken as the total error and is used in the fits. A
least $\chi^2$ fit method with~\cite{footnote}
$$
\chi^2 = \sum_{i=1}^{14}
         \frac{(\sigma^{\rm meas}_i-\sigma^{\rm fit}(m_i))^2}
              {(\Delta\sigma^{\rm meas}_i)^2}
$$
is used, where $\sigma^{\rm meas}_i\pm \Delta\sigma^{\rm meas}_i$
is the experimental measurement, and $\sigma^{\rm fit}(m_i)$ is
the cross section calculated from the model below with the
parameters from the fit. Here $m_i$ is the energy corresponds to
the $i$th energy point.

As the line shape above 4.42~GeV is unknown, it is not clear
whether the large cross section at high energy will decrease or
not. The data are fitted with two different scenarios.

Assuming the cross section follows the three-body phase space and
there is a narrow resonance at around 4.2~GeV, the cross sections
are fitted with the coherent sum of two amplitudes, a constant and
a constant-width relativistic BW function, i.e.,
$$
 \sigma(m)=|{\rm const}\cdot \sqrt{\rho(m)} +
 e^{i\phi}BW(m)\sqrt{\rho(m)/\rho(M)}|^2,
$$
where $\rho(m)$ is the 3-body phase space factor,
$BW(m)=\frac{\sqrt{12\pi\Gamma_{\EE}\BR(\pp\hc)\Gamma_{\rm tot}}}
{m^2-M^2+iM\Gamma_{\rm tot}}$, is the BW function for a vector
state, with mass $M$, width $\Gamma_{\rm tot}$, electron partial
width $\Gamma_{\EE}$, and the branching fraction to $\pp\hc$,
$\BR(\pp\hc)$, keep in mind that from the fit one can only extract
the product $\Gamma_{\EE}\BR(\pp\hc)$. The constant term ${\rm
const}$ and the relative phase, $\phi$, between the two amplitudes
are also free parameters in the fit together with the resonant
parameters of the BW function.

\begin{figure}[htbp]
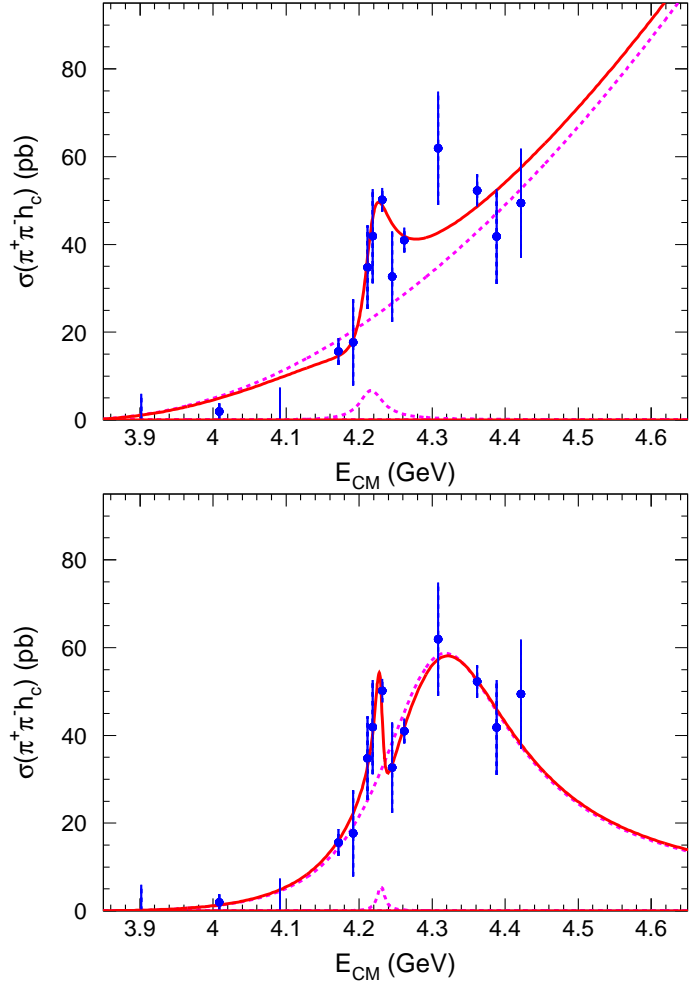

\centering
 \includegraphics[height=9.0cm,angle=-90]{BW+PS.epsi}\\
 \includegraphics[height=9.0cm,angle=-90]{2BW.epsi}
 \caption{The fit to the cross sections of $\EE\to \pp
h_c$ from BESIII and CLEO-c~(dots with error bars). Solid curves
show the best fits, and the dashed ones are individual component.
Top panel is the fit with the coherent sum of a phase space
amplitude and a BW function, and the bottom panel is the coherent
sum of two BW functions.}
  \label{fit_pipihc}
\end{figure}

The fit indicates the existence of a resonance (called $Y(4220)$
hereafter) with a mass of $(4216\pm 7)$~MeV/$c^2$ and width of
$(39\pm 17)$~MeV, and the goodness-of-the-fit is $\chi^2/{\rm ndf}
= 11/9$, corresponding to a confidence level of 27\%. There are
two solutions for the $\Gamma_{\EE}\times {\cal B}(Y(4220)\to
\pphc)$ which are $(3.2\pm 1.5)$~eV and $(6.0\pm 2.4)$~eV. Here
all the errors are from fit only. Fitting the cross sections
without the $Y(4220)$ results in a very bad fit, $\chi^2/{\rm ndf}
= 73/13$, corresponding to a confidence level of $2.5\times
10^{-10}$. The statistical significance of the $Y(4220)$ is
calculated to be $7.1\sigma$ comparing the two $\chi^2$s obtained
above and taking into account the change of the
number-of-degrees-of-freedom. Figure~\ref{fit_pipihc}~(top panel)
shows the final fit with the $Y(4220)$.

Assuming the cross section decreases at high energy, the cross
sections are fitted with the coherent sum of two constant-width
relativistic BW functions, i.e.,
$$
 \sigma(m) = |BW_1(m)\cdot \sqrt{\rho(m)/\rho(M_1)}
   + e^{i\phi}BW_2(m)\cdot \sqrt{\rho(m)/\rho(M_2)}|^2,
$$
where both $BW_1$ and $BW_2$ take the same form as $BW(m)$ above
but with different resonant parameters.

The fit indicates the existence of the $Y(4220)$ with a mass of
$(4230\pm 10)$~MeV/$c^2$ and width of $(12\pm 36)$~MeV, as well as
a broad resonance, the $Y(4290)$, with a mass of $(4293\pm
9)$~MeV/$c^2$ and width of $(222\pm 67)$~MeV. The
goodness-of-the-fit is $\chi^2/{\rm ndf} = 2/7$, corresponding to
a confidence level of 97\%, an almost perfect fit. There are two
solutions for the $\Gamma_{\EE}\times {\cal B}[Y(4220)/Y(4290)\to
\pphc]$ which are $(0.07\pm 0.07)~{\rm eV}/(16.1\pm 2.2)$~eV and
$(2.7\pm 4.9)~{\rm eV}/(19.0\pm 5.9)$~eV. Again, here the errors
are from fit only. Fitting the cross sections without the
$Y(4220)$ results in a much worse fit, $\chi^2/{\rm ndf} = 31/11$,
corresponding to a confidence level of $1.3\times 10^{-3}$. The
statistical significance of the $Y(4220)$ is calculated to be
$4.5\sigma$ comparing the two $\chi^2$s obtained above and taking
into account the change of the number-of-degrees-of-freedom.
Figure~\ref{fit_pipihc}~(bottom panel) shows the final fit with
the $Y(4220)$ and $Y(4290)$.

From the two fits showed above, we conclude that very likely there
is a narrow structure at around 4.22~GeV/$c^2$, although we are
not sure if there is a broad resonance at 4.29~GeV/$c^2$. We try
to average the results from the fits to give the best estimation
of the resonant parameters. For the $Y(4220)$, we obtain
\begin{eqnarray*}
M(Y(4220))                &=& (4216\pm 18)~\hbox{MeV}/c^2, \\
\Gamma_{\rm tot}(Y(4220)) &=& (39\pm 32)~\hbox{MeV}, \\
\Gamma^{Y(4220)}_{\EE}\times \BR[Y(4220)\to \pphc]
                          &=& (4.6\pm 4.6)~\hbox{eV}.
\end{eqnarray*}
While for the $Y(4290)$, we obtain
\begin{eqnarray*}
M(Y(4290))                &=& (4293\pm 9)~\hbox{MeV}/c^2, \\
\Gamma_{\rm tot}(Y(4290)) &=& (222\pm 67)~\hbox{MeV}, \\
\Gamma^{Y(4290)}_{\EE}\times \BR[Y(4290)\to \pphc]
                          &=& (18\pm 8)~\hbox{eV}.
\end{eqnarray*}
Here the errors include both statistical and systematic errors.
The results from the two solutions and the two fit scenarios are
covered by enlarged errors, the common systematic error in the
cross section measurement is included in the error of the
$\Gamma_{\EE}$.

It is noticed that the uncertainties of the resonant parameters of
the $Y(4220)$ are large, this is due to (1) the lack of data at CM
energies above 4.42~GeV which may discriminate which of the two
above scenarios is correct, and (2) the lack of high precision
measurements around the $Y(4220)$ peak, especially between 4.23
and 4.26~GeV. The two-fold ambiguity in the fits is a nature
consequence of the coherent sum of two amplitudes~\cite{zhuk},
although high precision data will not resolve the problem, they
will reduce the errors in $\Gamma_{\EE}$ from the above fits. As
the fit with a phase space amplitude predicts rapidly increasing
cross section at high energy, it is very unlikely to be true, so
the results from the fit with two resonances is more likely to be
true. More measurements from the BESIII experiments at CM energies
above 4.42~GeV and more precise data at around the $Y(4220)$ peak
will also be crucial to settle down all these problems.

There are thresholds of $D\bar{D}_1$~\cite{zhaoq1},
$\omega\chi_{cJ}$~\cite{zhenghq,yuancz},
$D_s^{\ast+}D_s^{\ast-}$~\cite{pdg} at the $Y(4220)$ mass region,
these make the identification of the nature of this structure very
complicated. The fits described in this paper supply only one
possibility of interpreting the data. In Ref.~\refcite{zhaoq2},
the BESIII measurements~\cite{zc4020} were described with the
presence of one relative $S$-wave $D \bar{D}_1 + c.c.$ molecular
state $Y(4260)$ and a non-resonant background term; while in
Ref.~\refcite{voloshin}, the BESIII data~\cite{zc4020} were fitted
with a model where the $Y(4260)$ and $Y(4360)$ are interpreted as
the mixture of two hadroncharmonium states. It is worth to point
out that various QCD calculations indicate that the
charmonium-hybrid lies in the mass region of these two $Y$
states~\cite{ccg_lqcd} and the $c\bar{c}$ tend to be in a
spin-singlet state. Such a state may couple to a spin-singlet
charmonium state such as $h_c$ strongly, this makes the $Y(4220)$
and/or $Y(4290)$ good candidates for the charmonium-hybrid states.

\section{Observation of charged charmoniumlike states}
\label{seczc}

Searching for the charged charmoniumlike state is the most
promising way of studying the exotic hadrons, since such a state
must contain at least four quarks and thus could not be a
conventional meson.

The Belle collaboration first reported evidence for a narrow
$Z(4430)^-$ peak, with mass $M=(4433\pm 4\pm 2)$~MeV/$c^2$ and
width $\Gamma=45^{+18+30}_{-13-13}$~MeV, in the $\pi^-\psip$
invariant mass distribution in $B\to K \pi^-\psip$
decays~\cite{Choi:2007wga}, and very soon reported another two
exotic $\pi^-\chi_{c1,2}$ structures in $B\to K \pi^-\chi_{c1}$
decays~\cite{zc12} at masses 4050 and 4250~MeV/$c^2$. The BaBar
collaboration did the same
analyses~\cite{Aubert:2008aa,Lees:2011ik}, but didnot confirm the
existence of these structures. On the other hand, the BaBar's
results didnot contradict the Belle evidence for these states due
to low statistics. This has been an open question for a very long
time since there were no new data available until very recently.

In the study of $\EE\to \ppjpsi$ at CM energies around 4.26~GeV,
the BESIII~\cite{zc3900} and Belle~\cite{belley_new} experiments
observed a charged charmoniumlike state, the $\zc$ in its
$\pi\jpsi$ decays, which was confirmed shortly after with CLEO
data at a CM energy of 4.17~GeV~\cite{seth_zc}. More recently,
BESIII observed a charged $Z_{c}(3885)$ state in $\EE\to
\pi^\pm(D\bar{D}^*)^\mp$~\cite{zc3885}, a charged $Z_{c}(4025)$
state in $\EE\to \pi^\pm(D^*\bar{D}^*)^\mp$~\cite{zc4025}, and a
charged $Z_{c}(4020)$ state in $\EE\to \pi^\pm(\pi^\mp
h_c)$~\cite{zc4020}. These states seem to indicate that a new
class of hadrons has been observed.

To take into account the interference effect between the
$Z(4430)^-$ and the $K^*$ intermediate states in $B\to K
\pi^-\psip$ decays, the Belle collaboration updated their
$Z(4430)^-$ results with a four-dimensional (4D) amplitude
analysis~\cite{zc4430pwa}. The $Z(4430)^-$ is observed with a
significance of $5.2\sigma$, a much larger mass of $M[Z(4430)^-] =
(4485\pm 22^{+28}_{-11})$~MeV/$c^2$, and a large width of
$\Gamma[Z(4430)^-] = (200^{+41+26}_{-46-35})$~MeV. The product
branching fractions are measured to be \( \BR(B^0\to
Z(4430)^-K^+)\times \BR(Z(4430)^-\to \pi^-\psip) =
(6.0^{+1.7+2.5}_{-2.0-1.4})\times 10^{-5}\), and spin-parity
$J^P=1^+$ is favored over the other assignments by more than
$3.4\sigma$. This was confirmed recently by the LHCb experiment in
a 4D model-dependent amplitude fit to a sample of $25176\pm 174$
$B^0\to K^+ \pi^-\psip$, $\psip\to \MM$ events~\cite{LHCbzc4430}.

Belle observed a $Z(4200)^-$ with more than $7.2\sigma$
significance and the evidence for the $Z(4430)^-$ in the
$\pi^\pm\jpsi$ invariant mass distribution in $B\to K \pi^-\jpsi$
decays~\cite{shencp_CPS_talk}.

As there are at least four quarks within the all these $Z_c^-$
states, they have been interpreted either as tetraquark states
with a pair of charm-anticharm quarks and a pair of light quarks,
molecular states of two charmed mesons ($\bar{D}D^*$,
$\bar{D^*}D^*$, $\bar{D}D_1$, $\bar{D^*}D_1$, etc.),
hadro-quarkonium states, or other configurations.

\subsection{\boldmath Observation of the $Z_c(3900)$ and $Z_c(3885)$}

\begin{figure}[htbp]
 \includegraphics[height=4.3cm]{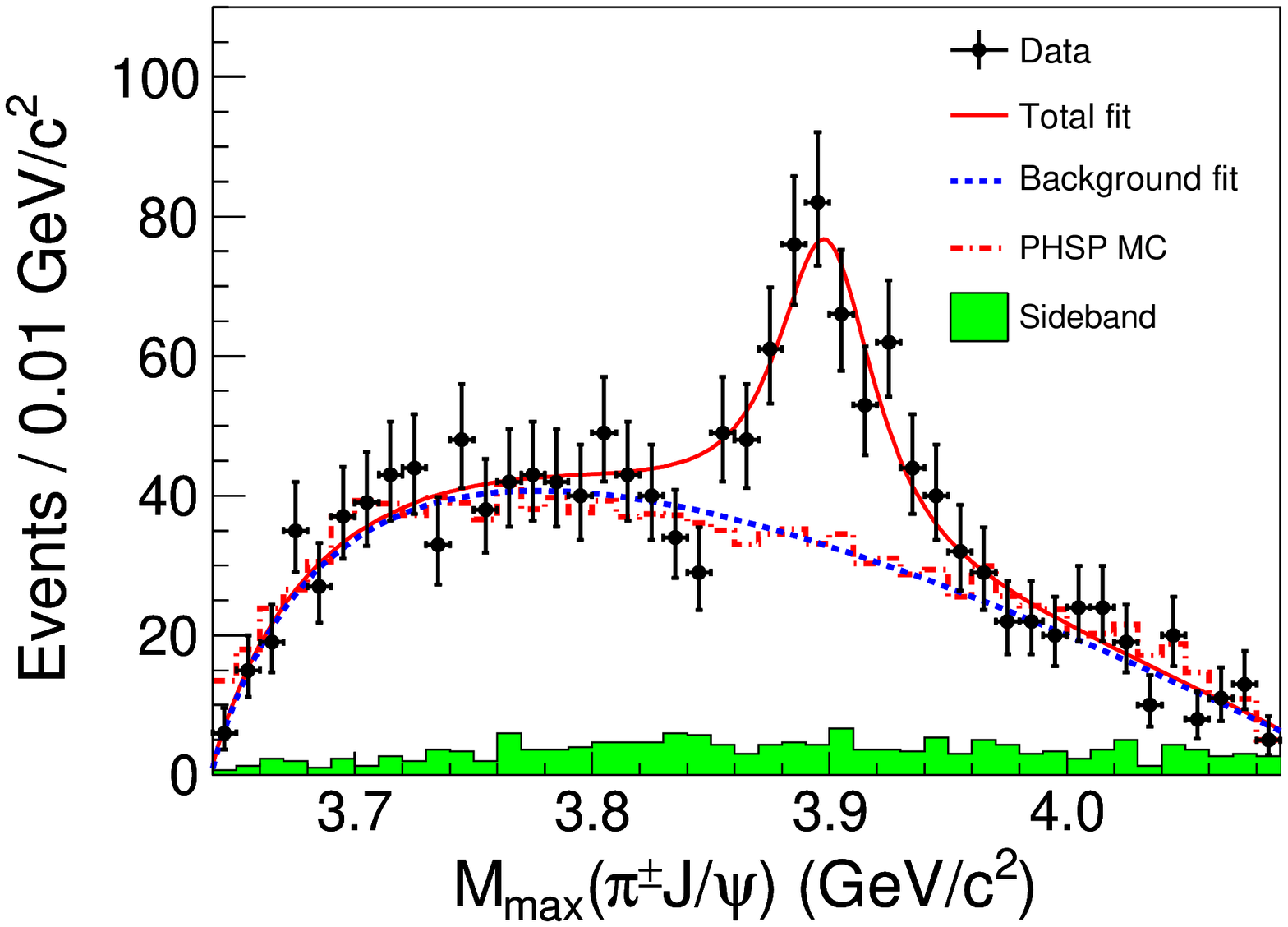}
 \includegraphics[height=4.3cm]{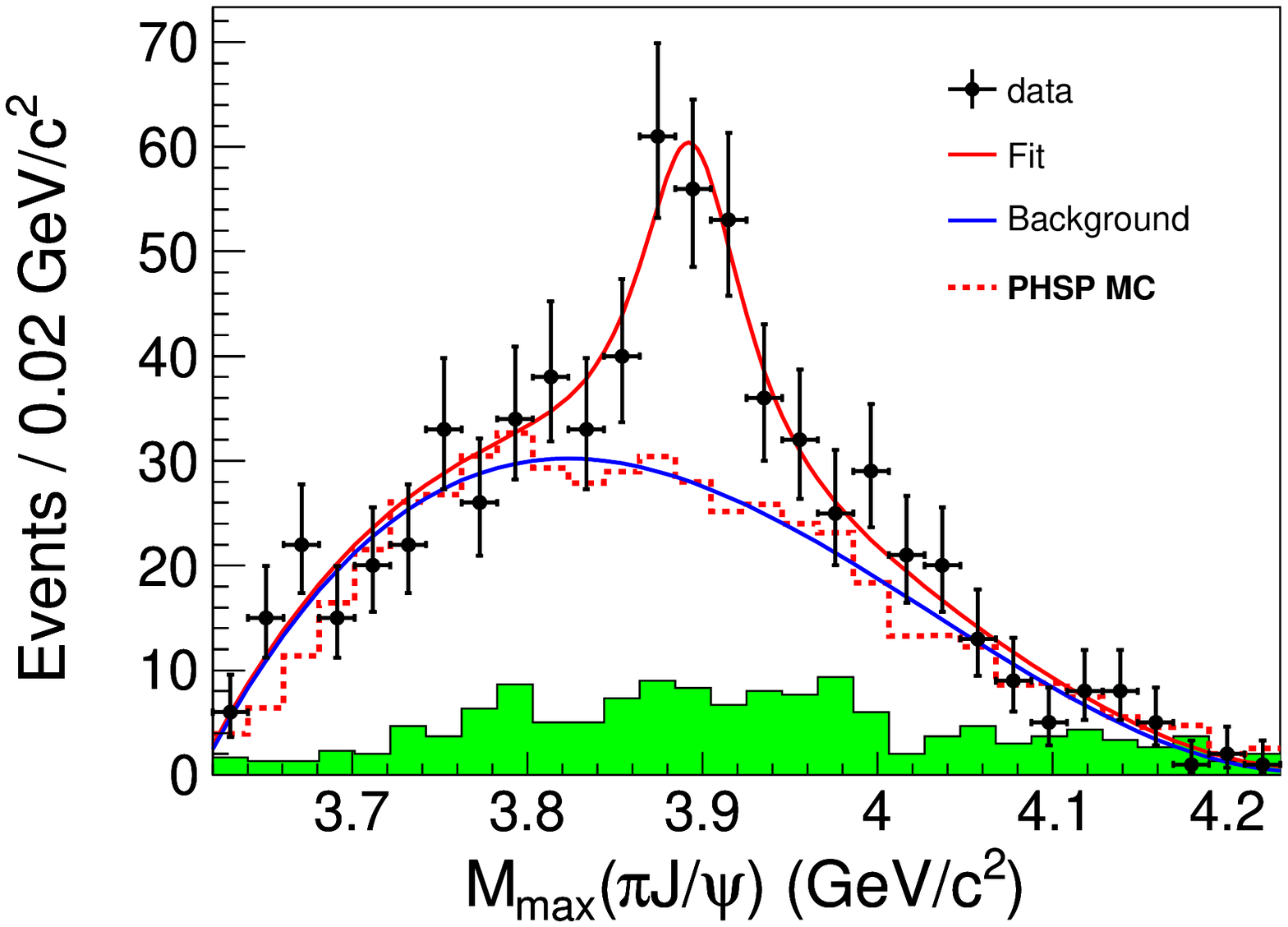}
 \caption{Unbinned maximum likelihood fit to the distribution of
the $M_{\mathrm{max}}(\pi J/\psi)$ (left panel from BESIII and
right panel from Belle). Points with error bars are data, the
curves are the best fit, the dashed histograms are the phase space
distributions and the shaded histograms are the non-$\ppjpsi$
background estimated from the normalized $\jpsi$ sidebands.}
\label{projfit}
\end{figure}

BESIII experiment studied the process $\EE\to \ppjpsi$ at a CM
energy of $4.260$~GeV using a 525~pb$^{-1}$ data
sample~\cite{zc3900}. A structure at around 3.9~GeV/$c^2$ is
observed in the $\pi^\pm \jpsi$ mass spectrum with a statistical
significance larger than $8\sigma$, which is referred to as the
$\zc$. A fit to the $\pi^\pm\jpsi$ invariant mass spectrum (see
Fig.~\ref{projfit}), neglecting interference, results in a mass of
$(3899.0\pm 3.6\pm 4.9)~{\rm MeV}/c^2$ and a width of $(46\pm
10\pm 20)$~MeV. Its production ratio is measured to be
$R=\frac{\sigma(\EE\to \pi^\pm \zc^\mp\to \ppjpsi))}
{\sigma(\EE\to \ppjpsi)}=(21.5\pm 3.3\pm 7.5)\%$.

At Belle experiment, the cross section of $\EE\to \ppjpsi$ is
measured from 3.8~GeV to 5.5~GeV using ISR method. The $\y$
resonance is observed and its resonant parameters are determined.
The intermediate states in $\y\to \ppjpsi$ decays are also
investigated~\cite{belley_new}. The $\zc$ (was named $Z(3900)^+$
in the Belle paper) state with a mass of $(3894.5\pm 6.6\pm
4.5)~{\rm MeV}/c^2$ and a width of $(63\pm 24\pm 26)$~MeV is
observed in the $\pi^\pm\jpsi$ mass spectrum (see
Fig.~\ref{projfit}) with a statistical significance larger than
$5.2\sigma$.

%%%% M(max) fit %%%%

The $\zc$ was confirmed shortly after with CLEO-c data at a CM
energy of 4.17~GeV~\cite{seth_zc}, the mass and width agree with
the BESIII and Belle measurements very well.

\begin{figure}[htbp]
\centering
  \includegraphics[width=6.0cm]{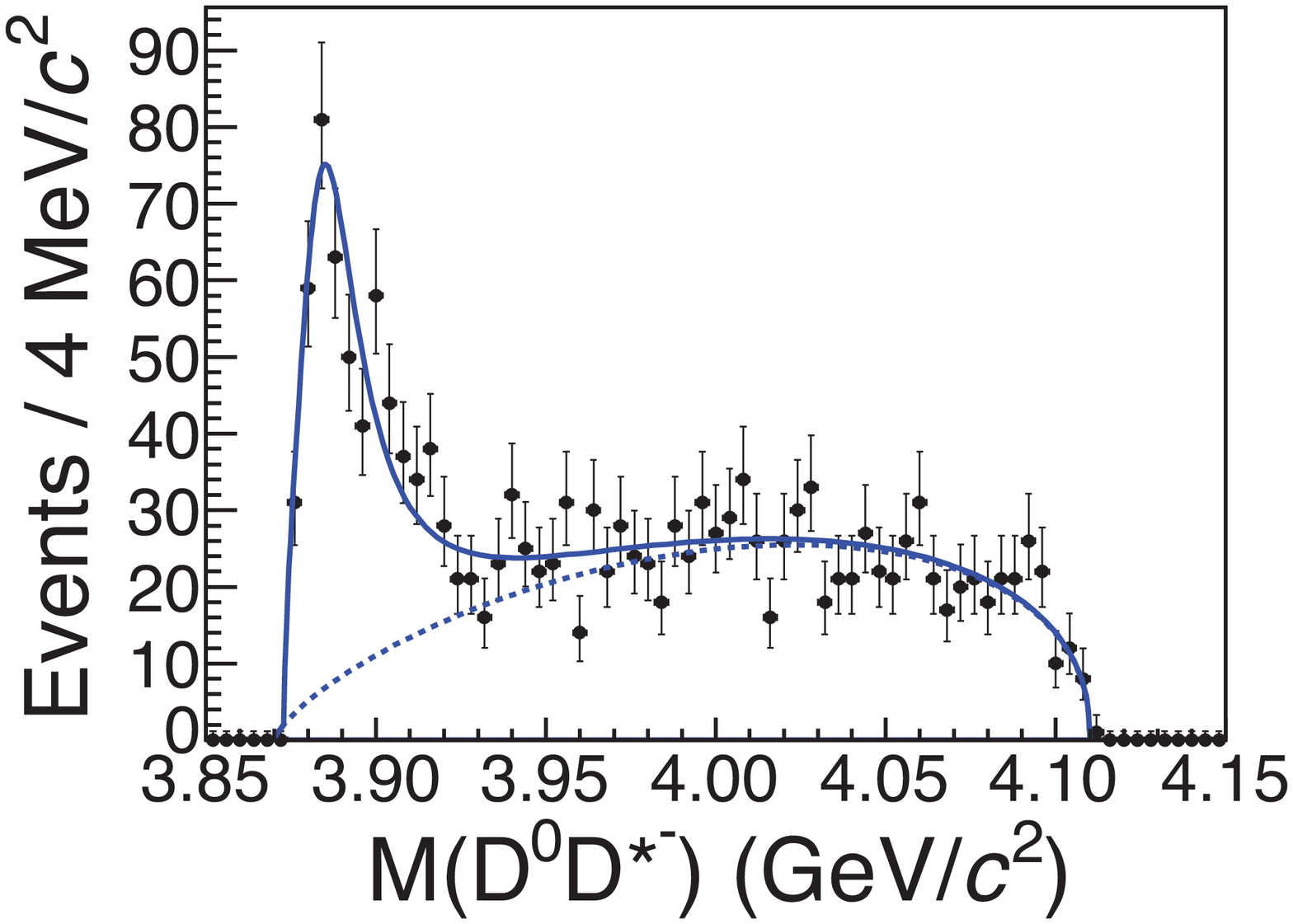}
  \includegraphics[width=6.0cm]{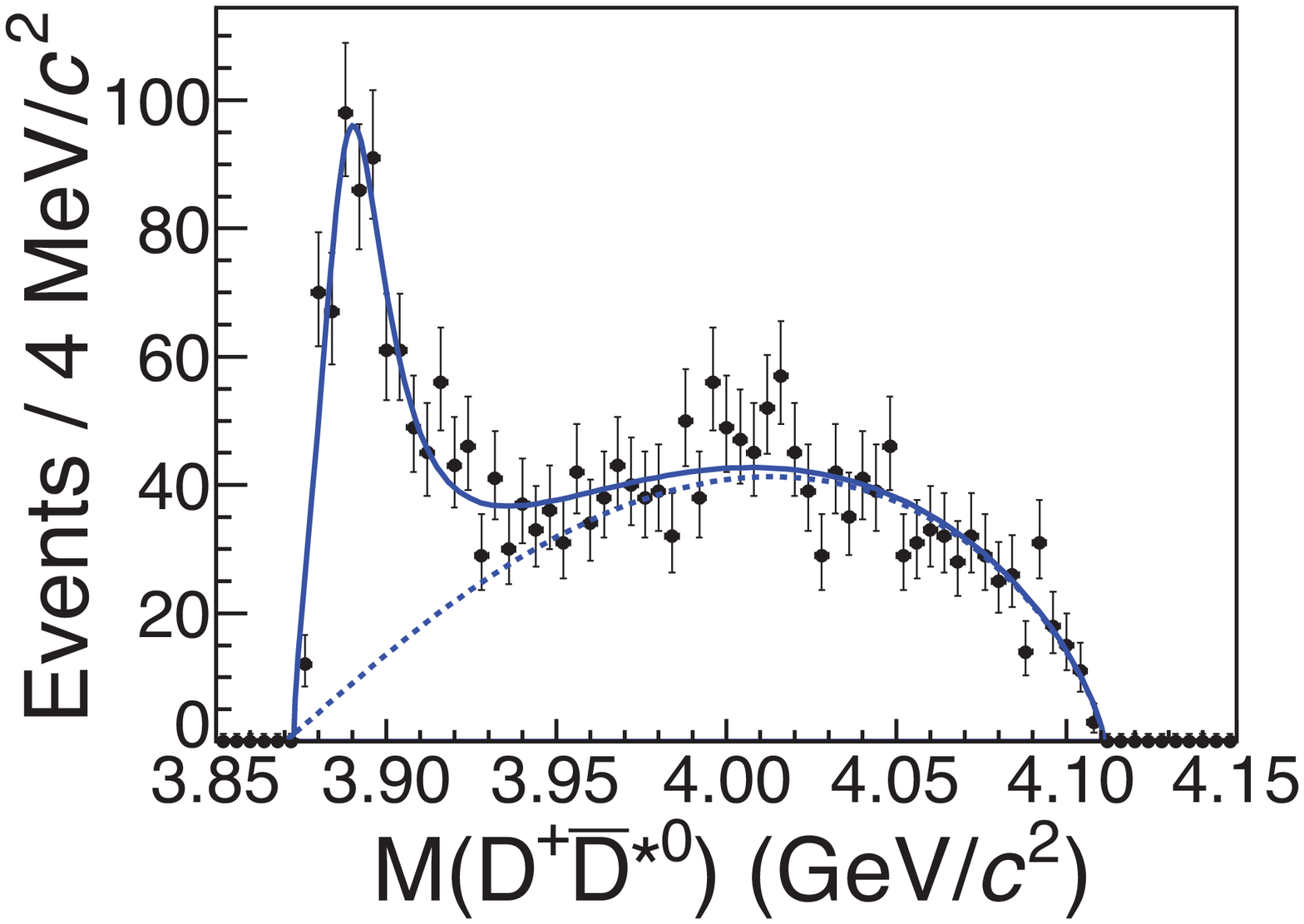}
\caption{The $M(D^0 D^{*-})$ (left) and $M(D^+\bar{D}^{*0})$
(right) distributions for selected events at $\sqrt{s}=4.26$~GeV.
The curves show the best fits. } \label{xuxp}
\end{figure}

This state is close to and above the $D\bar{D}^*$ mass threshold.
With the same data sample at $\sqrt{s}=4.26$~GeV, BESIII
experiment reported on a study of the process $\EE\to \pi^\pm
(D\bar{D}^*)^{\mp}$. A structure (referred to as $Z_c(3885)$) is
observed in the $(D\bar{D}^*)^{\mp}$ invariant mass
distribution~\cite{zc3885}. When fitted to a mass-dependent-width
BW function, the pole mass and width are determined to be $(3883.9
\pm 1.5 \pm 4.2)$~MeV/$c^2$ and $(24.8\pm 3.3 \pm 11.0)$~MeV,
respectively (see Fig.~\ref{xuxp}). The angular distribution of
the $Z_c(3885)$ system favors a $J^{P}=1^{+}$ assignment for the
structure and disfavors $1^-$ or $0^-$. The production rate is
measured to be $\sigma(\EE \to \pi^{\pm} Z_c(3885)^{\mp})\times
\BR(Z_c(3885)^{\mp}\to (D\bar{D}^*)^{\mp}) =(83.5\pm 6.6 \pm
22.0)$~pb.

An important question is whether or not the $Z_c(3885)$ is the
same as the $\zc$. The mass and width of the $Z_c(3885)$ are
$2\sigma$ and $1\sigma$, respectively, below those of the
$Z_c(3900)$ observed by BESIII and Belle. However neither fit
considers the possibility of interference with a coherent
non-resonant background that could shift the results. A $J^P$
quantum number determination of the $Z_c(3900)$ would provide an
additional test of this possibility.

Assuming the $Z_c(3885)$ structure is due to the $Z_c(3900)$, one
obtains $\frac{\Gamma(Z_c(3885)\to D\bar{D}^*)}
{\Gamma(Z_c(3900)\to \pi\jpsi)} = 6.2\pm 1.1\pm 2.7$. This ratio
is much smaller than typical values for decays of conventional
charmonium states above the open charm threshold. For example:
$\Gamma (\psi(3770)\to D\bar{D})/\Gamma(\psi(3770)\to
\pp\jpsi)=482\pm 84$~\cite{pdg} and $\Gamma (\psi(4040)\to
D^{(*)}\bar{D}^{(*)}) / \Gamma(\psi(4040)\to \eta\jpsi)=192\pm
27$~\cite{etajpsi}. This suggests the influence of very different
dynamics in the $\y$-$\zc$ system.

Assuming that the $Z_c(3885)$ and $Z_c(3900)$ are the same
structure and neglecting the slightly different parametrization of
the resonances, a weighted average of the above four measurements
yields the best estimation of the resonant parameters of the
$\zc$, i.e., $M_{\zc}=(3888.6\pm 2.7)$~MeV/$c^2$ and
$\Gamma_{\zc}=(34.7\pm 6.6)$~MeV.

\subsection{\boldmath Observation of the $Z_c(4020)$ and $Z_c(4025)$}

BESIII measured $\EE\to \pphc$ cross sections~\cite{zc4020} at CM
energies between 3.90 and 4.42~GeV as described in
Sec.~\ref{sec_y}.

\begin{figure}[htbp]
\begin{center}
\includegraphics[width=10cm]{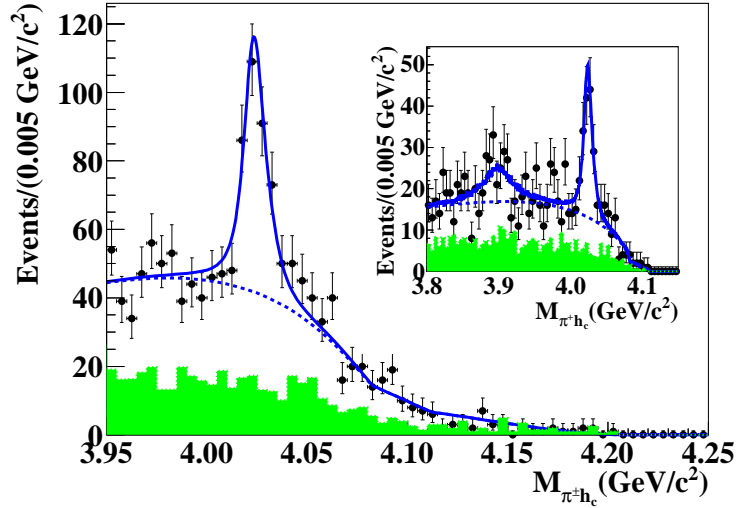}
\caption{Sum of the simultaneous fits to the $M(\pi^\pm h_c)$
distributions at 4.23~GeV, 4.26~GeV, and 4.36~GeV in the BESIII
data; the inset shows the sum of the simultaneous fit to the
$M_{\pi^+ h_c}$ distributions at 4.23~GeV and 4.26~GeV with $\zc$
and $\zcp$. Dots with error bars are data; shaded histograms are
normalized sideband background; the solid curves show the total
fit, and the dotted curves the backgrounds from the fit.}
\label{1Dfit}
\end{center}
\end{figure}

Intermediate states are studied by examining the Dalitz plot of
the selected $\pphc$ candidate events. The $\hc$ signal is
selected using $3.518 < M_{\gamma \eta_c} < 3.538$~GeV/$c^2$,
$\pphc$ samples of 859 events at 4.23~GeV, 586 events at 4.26~GeV,
and 469 events at 4.36~GeV are obtained with purities of ~65\%.
While there are no clear structures in the $\pp$ system, there is
clear evidence for an exotic charmoniumlike structure in the
$\pi^\pm\hc$ system as clearly shown in the Dalitz plot.
Figure~\ref{1Dfit} shows the projection of the $M(\pi^\pm\hc)$
(two entries per event) distribution for the signal events, as
well as the background events estimated from normalized $\hc$ mass
sidebands. There is a significant peak at around 4.02~GeV/$c^2$
(the $\zcp$), and there are also some events at around
3.9~GeV/$c^2$ (inset of Fig.~\ref{1Dfit}), which could be the
$\zc$. The individual data sets at 4.23~GeV, 4.26~GeV and 4.36~GeV
show similar structures.

An unbinned maximum likelihood fit is applied to the
$M(\pi^\pm\hc)$ distribution summed over the 16 $\eta_c$ decay
modes. The data at 4.23~GeV, 4.26~GeV, and 4.36~GeV are fitted
simultaneously with the same signal function with common mass and
width. Figure~\ref{1Dfit} shows the fit results. The mass of the
$\zcp$ is measured to be $(4022.9\pm 0.8\pm 2.7)~{\rm MeV}/c^2$,
and the width is $(7.9\pm 2.7\pm 2.6)$~MeV. The statistical
significance of the $\zcp$ signal is calculated by comparing the
fit likelihoods with and without the signal. Besides the nominal
fit, the fit is also performed by changing the fit range, the
signal shape, or the background shape. In all cases, the
significance is found to be greater than $8.9\sigma$.

Adding a $\zc$ with mass and width fixed to the BESIII
measurement~\cite{zc3900} in the fit, results in a statistical
significance of 2.1$\sigma$ (see the inset of Fig.~\ref{1Dfit}).
The upper limits on the production cross sections are set as
$\sigma(\EE\to \pi^\pm \zc^\mp\to \pphc) <13$~pb at 4.23~GeV and
$<11$~pb at 4.26~GeV, at the 90\% confidence level (C.L.). This is
lower than that of $\zc\to \pi^\pm\jpsi$~\cite{zc3900}.

\begin{figure}[htbp]
\centering
\includegraphics[width=8cm]{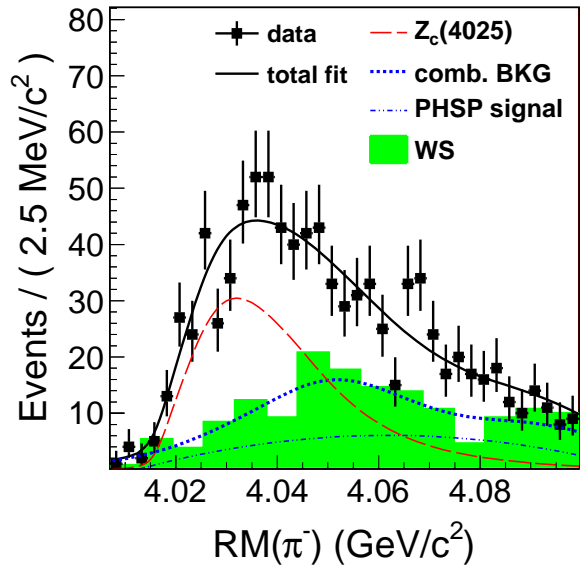}
\caption{Unbinned maximum likelihood fit to the $\pi^\mp$ recoil
mass spectrum in $e^+e^- \to (D^{*} \bar{D}^{*})^{\pm} \pi^\mp$ at
$\sqrt{s}=4.26$~GeV at BESIII. } \label{fig:fit}
\end{figure}

BESIII experiment also studied the process $e^+e^- \to (D^{*}
\bar{D}^{*})^{\pm} \pi^\mp$ at a CM energy of 4.26~GeV using a
827~pb$^{-1}$ data sample~\cite{zc4025}. Based on a partial
reconstruction technique, the Born cross section is measured to be
$(137\pm 9\pm 15)$~pb. A structure near the $(D^{*}
\bar{D}^{*})^{\pm}$ threshold in the $\pi^\mp$ recoil mass
spectrum is observed, which is denoted as the $Z_c(4025)$ (see
Fig.~\ref{fig:fit}. The measured mass and width of the structure
are $(4026.3\pm 2.6\pm 3.7)$~MeV/$c^{2}$ and $(24.8\pm 5.6\pm
7.7)$~MeV, respectively, from a fit with a constant-width BW
function for the signal. Its production ratio
$\frac{\sigma(e^+e^-\to Z^{\pm}_c(4025)\pi^\mp \to (D^{*}
\bar{D}^{*})^{\pm} \pi^\mp)}{\sigma(e^+e^-\to (D^{*}
\bar{D}^{*})^{\pm} \pi^\mp)}$ is determined to be $0.65\pm 0.09\pm
0.06$.

The $Z_c(4025)$ parameters agree within 1.5$\sigma$ of those of
the $\zcp$. Very probably they are the same state. As the results
on the $Z_c(4025)$ is only from the data at 4.26~GeV, extending
the analysis to the data at 4.23~GeV and 4.36~GeV will probably
give us a definite answer.

\subsection{\boldmath Confirmation of the $Z(4430)^-$}

Observed by the Belle collaboration~\cite{Choi:2007wga} but not
confirmed by the BaBar collaboration~\cite{Aubert:2008aa}, the
existence of the $Z(4430)^-$ has been questioned until very
recently the LHCb experiment~\cite{LHCbzc4430} reported a 4D
model-dependent amplitude fit to a much larger sample of $B^0\to
K^+ \pi^-\psip$ events. In this LHCb analysis, the $Z(4430)^-$ was
observed at larger than 13.9$\sigma$ level, and the spin-parity is
determined to be $1^+$.

$25,176\pm 174$ $B^0\to\psip \pi^-K^+$, $\psip\to \mu^+\mu^-$
events are selected for the 4D amplitude analysis at
LHCb~\cite{LHCbzc4430}. The $B^0$ candidates are selected using
particle identification information, transverse momentum
thresholds and requiring separation of the tracks and of the $B^0$
vertex from the primary $pp$ interaction points. The background
fraction is determined from the $B^0$ candidate invariant mass
distribution to be $(4.1\pm 0.1)\%$. The background is dominated
by combinations of $\psip$ from $B$ decays with random kaons and
pions.

The LHCb amplitude model includes all known $K^{*0}\to K^+\pi^-$
resonances with nominal mass within or slightly above the
kinematic limit in $B^0\to K^+ \pi^-\psip$ decays: $K^*_0(800)$,
$K^*_0(1430)$ for $J=0$; $K^*(892)$, $K^*(1410)$ and $K^*(1680)$
for $J=1$; $K^*_2(1430)$ for $J=2$; and $K^*_3(1780)$ for $J=3$.
They also include a non-resonant $J=0$ term in the fits.

The above model does not describe the data well, the confidence
level of the fit is below $2\times 10^{-6}$, equivalent to
$4.8\sigma$ in the Gaussian distribution, according to toy MC
simulation. By adding a $Z(4430)^-$ component with $J^P=1^+$ in
the $\pi\psip$ system to the amplitude, the C.L. of the fit
improves to a few percent level, indicating a reasonable fit. The
resonant parameters are found to be $M[Z(4430)^-]=(4475\pm
7_{-25}^{+15})$~MeV/$c^2$, $\Gamma[Z(4430)^-]=(172\pm
13_{-34}^{+37})$~MeV. The spin-parity is found to be $1^+$, the
$0^-$, $1^-$, $2^+$ and $2^-$ hypotheses are ruled out at very
high significance level.

With the large statistics, LHCb experiment is able to divide the
$M^2[\pi\psip]$ bins around the $Z(4430)^-$ peak and fit the real
and imaginary parts of the $Z(4430)^-$ amplitudes. The resulting
Argand diagram is consistent with a rapid change of the
$Z(4430)^-$ phase when its magnitude reaches the maximum, a
behavior characteristic of a resonance.

With the LHCb results, the first charged charmoniumlike state
$Z(4430)^-$ is established after seven years of its discovery at
Belle~\cite{zc4430pwa}.

\subsection{\boldmath Observation of the $Z(4200)^-$}

The Belle experiment analyzed $B^0\to K^- \pi^+ \jpsi$ with
$\jpsi\to \EE$ or $MM$ with the full $\Upsilon(4S)$ data sample,
corresponds to 711~fb$^{-1}$ data with 772 million $B\bar{B}$
pairs~\cite{shencp_CPS_talk}. About 30,000 signal events are
selected and a 4D amplitude analysis is performed with
$K^*_0(800)$, $K^*_0(1430)$, and $K^*_0(1950)$ for $J=0$;
$K^*(892)$, $K^*(1410)$, and $K^*(1680)$ for $J=1$; $K^*_2(1430)$
and $K^*_2(1980)$ for $J=2$; $K^*_3(1780)$ for $J=3$, and
$K^*_4(2045)$ for $J=4$ in the $K\pi$ system; and a $J^P=1^+$ BW
function in the $\pi\jpsi$ system.

A resonant (named $Z_c(4200)$) is needed to describe the data with
a statistical significance of more than 7.2$\sigma$ including the
systematic effect. The resonant parameters are determined to be
$M[Z(4200)^-] = (4196_{-29-\phantom{1}6}^{+31+17})$~MeV/$c^2$,
$\Gamma[Z(4200)^-] = (370_{-70-85}^{+70+70})$~MeV. The product
branching fractions are measured to be
\[ \BR(B^0\to Z(4200)^-K^+)\times \BR(Z(4200)^-\to \pi^-\jpsi)
= (2.2^{+0.7+1.1}_{-0.5-0.6})\times 10^{-5}.\] The spin-parity is
found to be $1^+$, the $0^-$, $1^-$, $2^+$ and $2^-$ hypotheses
are ruled out at at least 5.6$\sigma$ level.

In addition, a strong evidence for $Z(4430)^-\to \pi^-\jpsi$ is
observed, with a significance level of more than $4.0\sigma$.
However, no significant $\zc$ is observed. The product branching
fractions are measured to be
\[ \BR(B^0\to Z(4430)^-K^+)\times \BR(Z(4430)^-\to \pi^-\jpsi)
= (5.4^{+4.0+1.1}_{-1.0-0.9})\times 10^{-6}.\] Comparing with the
measurement of $Z(4430)^-\to \pi^-\psip$ listed above, we found
that $\frac{\BR(Z(4430)^-\to \pi^-\jpsi)}{\BR(Z(4430)^-\to
\pi^-\psip)}=0.09^{+0.18}_{-0.05}$.

\subsection{Discussions}

In Table~\ref{zc_summary} we summarize all the charged
charmoniumlike states reported so far with the mass and width from
a simple weighted average of all the available measurements. While
the $\zc$, $\zcp$, and $Z(4430)^-$ are established, the two states
observed in $B\to K \chi_{c1}\pi^-$ and the newly observed
$Z(4200)^-$ need further confirmation.

\begin{table}[hbtp]
\tbl{Summary of the $Z_c$ states.}
 {\begin{tabular}{@{}lccl@{}} \toprule
  state       & mass (MeV/$c^2$) & width (MeV)    & comments  \\  \hline
  $\zc^-$       &  $3888.6\pm 2.7$ &  $34.7\pm 6.6$ & Weighted average of
     Refs.~\citen{zc3900,belley_new,seth_zc,zc3885} \\
  $\zcp^-$      &  $4023.9\pm 2.4$ & $10.2\pm 3.5$   & Weighted average of
     Refs.~\citen{zc4020,zc4025} \\
  $Z(4050)^-$ &  $4051{^{+24}_{-43}}$ & $82^{+51}_{-28}$
          & Ref.~\refcite{zc12}, need confirmation \\
  $Z(4200)^-$ &  $4196^{+35}_{-30}$ & $370^{+\phantom{1}99}_{-110}$
           & Ref.~\refcite{shencp_CPS_talk}, need confirmation \\
  $Z(4250)^-$ &  $4248^{+185}_{-\phantom{1}45}$ & $177^{+321}_{-\phantom{1}72}$
     & Ref.~\refcite{zc12}, need confirmation \\
  $Z(4430)^-$ &  $4478\pm 20$ & $181\pm 33$  &
      Weighted average of Refs.~\citen{zc4430pwa,LHCbzc4430} \\
  \botrule
\end{tabular}\label{zc_summary}}
\end{table}

The nature of these states have been discussed for a long time,
and there are many proposals~\cite{epjc-review} which will will
not repeat. It is obvious that some of these states are close to
open charm threshold such as $D\bar{D}^*$ [$\zc$], $D^*\bar{D}^*$
[$\zcp$], $D^*\bar{D}_1$ [$Z(4430)^-$], however, the other states
may not be very close to the thresholds.

The understanding of these $Z_c^-$ states and the similar states
in the $b\bar{b}$ system~\cite{belle_zb} may help in the
development of the QCD at non-perturbative domain.

\section{Summary and perspectives}

There are lots of charmoniumlike states observed in charmonium
mass region but many of them show properties different from the
naive expectation of conventional charmonium states. The BESIII
experiment is now producing results on these XYZ states. The
observation of the charged charmonium states, $\zc$, $\zcp$,
$Z(4430)^-$, and other states, may indicate one kind of the exotic
states has been observed.

In the near future, BESIII experiment~\cite{bes3} will accumulate
more data between 4.0 and 4.6~GeV for further study; the Belle II
experiment~\cite{belle2} under construction, with about
50~ab$^{-1}$ data accumulated, will surely improve our
understanding of all these states.

LHCb experiment~\cite{lhcb} has produced lots of interesting
results, with even more statistics expected, it may further
contribute to the XYZ particle study. An immediate effort would be
the study of $B\to K \pi^-\chi_{c1}$ and $B\to K \pi^-\jpsi$ for a
high statistics search for the three $Z_c$ states waiting for
confirmation.

PANDA~\cite{panda}, the $p\bar{p}$ annihilation experiment which
is designed to study the charmonium and charmoniumlike states,
will be able to contribute to the XYZ paricle study in a very
different way. With a very small beam energy spread, it may
measure the line shape of the $\xx$ and many other neutral XYZ
states; of course, the charged charmoniumlike states can also be
produced in company with a charged meson.

\section*{Acknowledgments}

We thank the organizers for the invitation. This work is supported
in part by National Natural Science Foundation of China (NSFC)
(10825524, 10935008, 11235011).

%\begin{thebibliography}{000} %for 3 digits


\begin{thebibliography}{00}
\bibitem{footnote} For the three low statistics energy points,
the $\chi^2$ is not well defined. We take the central values
listed in Table~\ref{scan-data} as nominal values, and vary the
central values and statistical errors in a wide range to estimate
the possible bias in this assumption. The bias is found to be
small and is considered as systematic error of the results.
\end{thebibliography}
\end{document}